\documentclass[journal=jacsat,manuscript=article]{achemso}
\setkeys{acs}{keywords = true,articletitle=true, maxauthors=100}
\usepackage[version=3]{mhchem} 

\author{G\"{o}ksel M{\i}s{\i}rl{\i}}
\affiliation{School of Computing and Mathetmatics, Keele University}
\email{g.misirli@keele.ac.uk}

\title[A comparative analysis for SARS-CoV-2]
  {A comparative analysis for SARS-CoV-2}

\abbreviations{IR,NMR,UV}

\begin{document}

\begin{abstract}
COVID-19 has affected the world tremendously. It is critical that biological experiments and clinical designs are informed by computational approaches for time- and cost-effective solutions. Comparative analyses particularly can play a key role to reveal structural changes in proteins due to mutations, which can lead to behavioural changes, such as the increased binding of the SARS-CoV-2 surface glycoprotein to human ACE2 receptors. The aim of this report is to provide an easy to follow tutorial for biologists and others without delving into different bioinformatics tools. More complex analyses such as the use of large-scale computational methods can then be utilised. Starting with a SARS-CoV-2 genome sequence, the report shows visualising DNA sequence features, deriving amino acid sequences, and aligning different genomes to analyse mutations and differences. The report provides further insights into how the SARS-CoV-2 surface glycoprotein mutated for higher binding affinity to human ACE2 receptors, compared to the SARS-CoV protein, by integrating existing 3D protein models.
\end{abstract}

\section{Introduction}
Severe acute respiratory syndrome (SARS)-related coronaviruses have previously caused two pandemics \cite{Zhou2020}. The recent SARS Coronavirus 2 (SARS2-CoV-2) outbreak has had unprecedented effects so far. It is essential to develop data integration mechanisms in order to gain insights using data that already exists. For example, genome-wide comparisons can be used to inform subsequent computational analyses which can potentially be used to search for drugs and to develop computational models.

Taxonomy classifications offer a systematic approach to link data from different organisms. The taxonomy id for SARS-CoV-2 is reported as 2697049 by the National Center for Biotechnology (NCBI) taxonomy browser\cite{NCBI-Taxonomy} which also lists synonyms that can be used when searching for information in different databases (\url{https://www.ncbi.nlm.nih.gov/Taxonomy/Browser/wwwtax.cgi?id=2697049}). Moreover, the taxonomy id 694009 is used to group all SARS-related coronavirus taxonomies. This list can especially be useful to search for related sequences in order to infer information via comparative genomics approaches. For example, the NCBI Virus \cite{NCBI-Virus} database can be queried with these taxonomy ids to retrieve SARS-CoV-2 related nucleotide and protein sequences. 


The SARS-CoV-2 genome sequences can also be accessed directly from the NCBI's GenBank database (\url{https://www.ncbi.nlm.nih.gov/genbank/sars-cov-2-seqs}). Related information includes where a virus was isolated from. Some of these GenBank files include only sequences and some of them provide annotations about genome locations of important sequence features.

This tutorial initially uses two genome sequences: one for SARS-CoV-2 and another one for SARS-CoV. Regarding the former, the GenBank entry LC528232 was selected. LC528232 was created for a strain which was isolated in Japan. Regarding the latter, the GenBank entry AP006557, which was deposited in 2006, was selected.

\section{Genomic features}

\begin{figure*}[!thb]
\centering
  \includegraphics[width=1\textwidth]{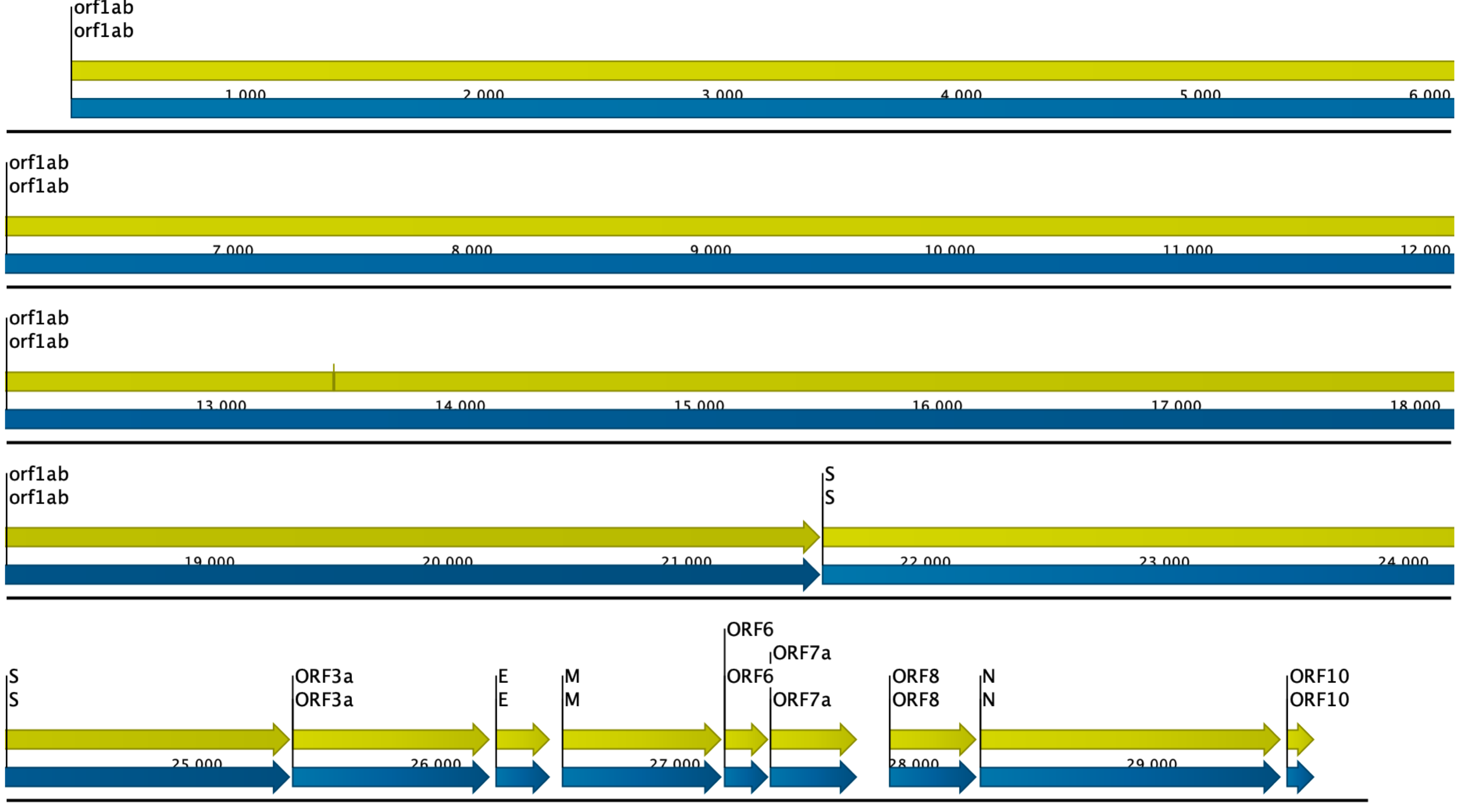}
    \caption{SARS-CoV-2 sequence features that are captured in GenBank files are shown.}
\label{fig:features}
\end{figure*}

The GenBank files include annotations about sequence features, which can be visualised and analysed using various tools. In this report, the analyses were carried out using CLC Genomics Workbench 20.0.3. GenBank files were initially imported into this tool which was then used to analyse the annotated sequence features in more detail. For example, the surface glycoprotein denoted as `S' is shown in Figure \ref{fig:features}.

\begin{figure*}[!thb]
\centering
  \includegraphics[width=1\textwidth]{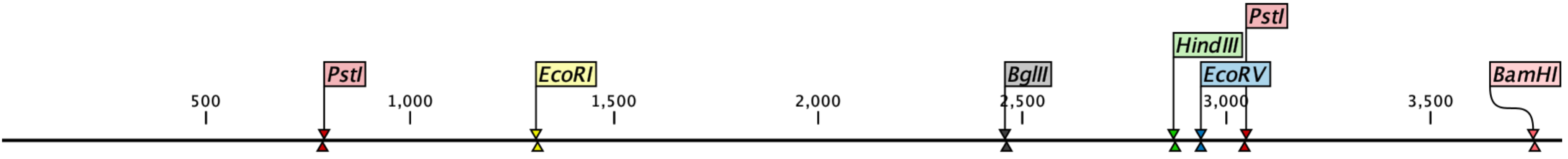}
    \caption{Restriction sites in the surface glycoprotein CDS, labelled as `S'.}
\label{fig:s_restriction_sites}
\end{figure*}

In order understand the mutations in the surface glycoprotein, a new entry for the related coding sequence (CDS) was created in CLC Genomics Workbench. The corresponding CDS feature was selected and the nucleotides were copied to create this new entry. CLC Genomics Workbench was then used to display restriction sites and protein translation information (Figure \ref{fig:s_restriction_sites}).

\section{Predicting secondary structures}
\begin{figure}[!thb]
\centering
  \includegraphics[width=1\textwidth]{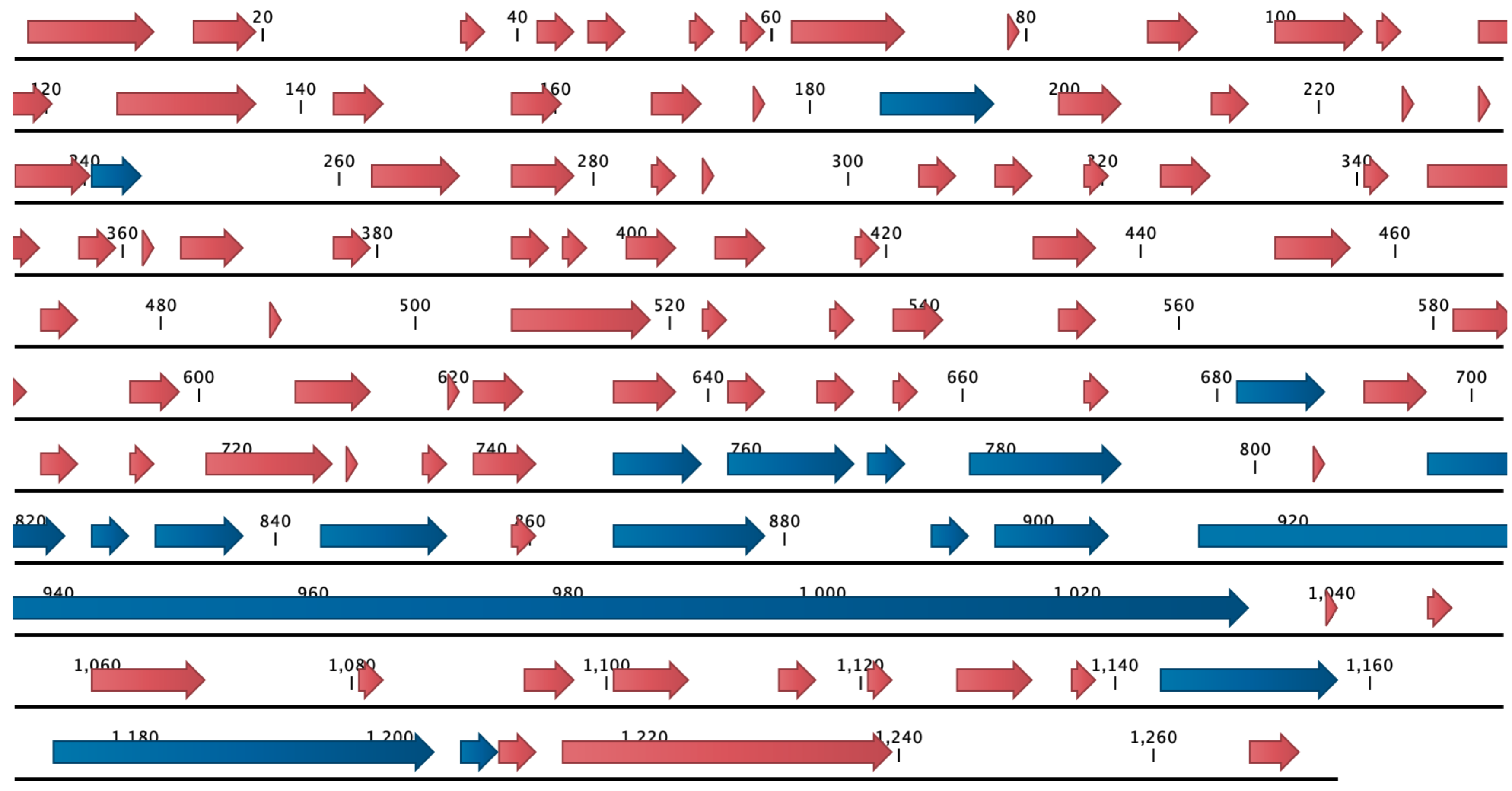}
    \caption{The surface glycoprotein secondary structure predictions. Red arrows represent beta-strands and blue arrows represent alpha-helices.}
\label{fig:s_secondary_structure_predictions}
\end{figure}

Understanding secondary structure formations and where mutations occur can provide insights into the effects of these mutations in the SARS-CoV-2 surface glycoprotein CDS. Hence, a new entry for the corresponding amino acid sequences was created in CLC Genomics Workbench by using the `\texttt{Toolbox - Classical Sequence Analysis - Nucleotide \\Analysis - Translate to Protein}' option. Alpha-helix and beta-strand formations were predicted using the `\texttt{Toolbox - Classical Sequence Analysis - Protein Analysis - Predict Secondary Structure}' option. The inferred information was then incorporated as annotations into the entry for the amino acid sequences (Figure \ref{fig:s_secondary_structure_predictions}).

\subsection{Sequence alignment}
\begin{figure}[!thb]
\centering
  \includegraphics[width=1\textwidth]{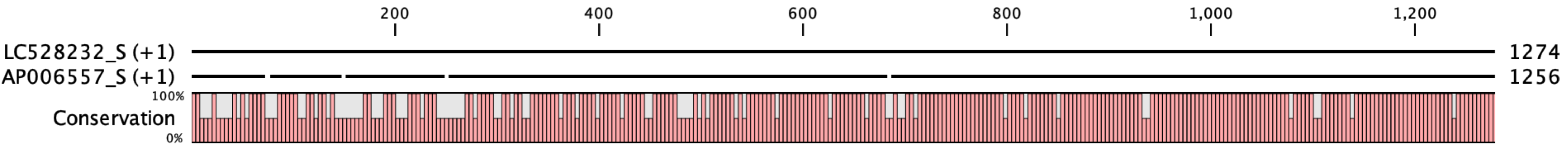}
    \caption{Aligned surface glycoprotein amino acid sequences from SARS-CoV-2 (LC528232) and SARS-CoV (AP006557).}
\label{fig:s_alignment_1}
\end{figure}

It is reported that the  SARS-CoV-2 surface glycoprotein is optimised to bind to human ACE2 receptors \cite{Andersen2020}. In order to analyse the effects of these mutations further, the protein sequence can be aligned to previously known similar sequences. Figure \ref{fig:s_alignment_1} shows the alignment of the surface glycoprotein amino acid sequences from SARS-CoV-2 (LC528232) and SARS-CoV (AP006557). Although the alignment shows high similarities between the amino acid sequences, gaps and mutations also exist.

\begin{figure}[!thb]
\centering
  \includegraphics[width=1\textwidth]{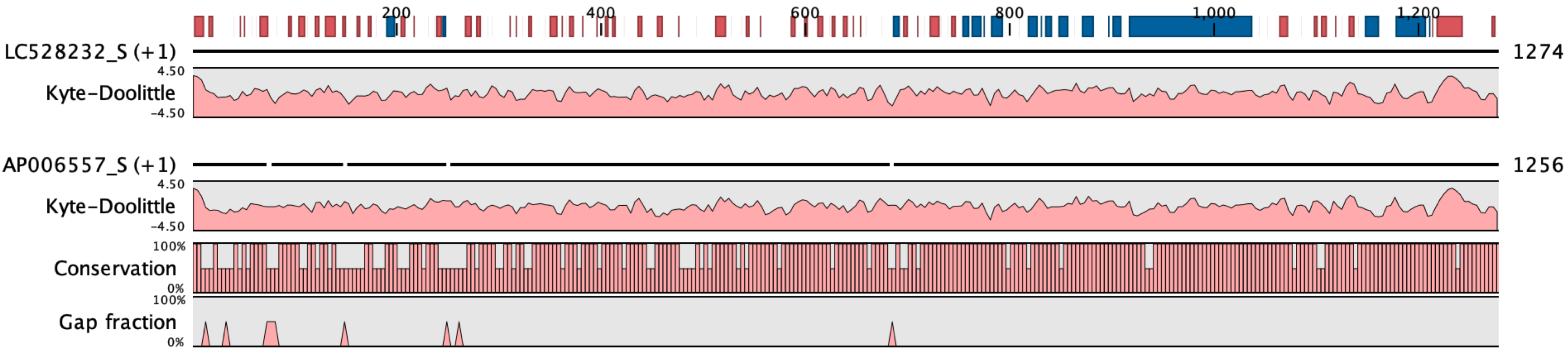}
    \caption{Aligned surface glycoprotein amino acid sequences from SARS-CoV-2 (LC528232) and SARS-CoV (AP006557). Red blocks at the top represent beta-strands and blue blocks represent alpha-helices.}
\label{fig:s_alignment_2}
\end{figure}

A more detailed view of the alignment of the two protein sequences can be seen in Figure \ref{fig:s_alignment_2}. Secondary structures are integrated into the view. The surface exposed regions are compared using different options such as the Kyte-Doolittle scale.  Additional options can be configured from the `\texttt{Alignment Settings - Protein info}' tab.

\begin{figure}[!thb]
\centering
  \includegraphics[width=1\textwidth]{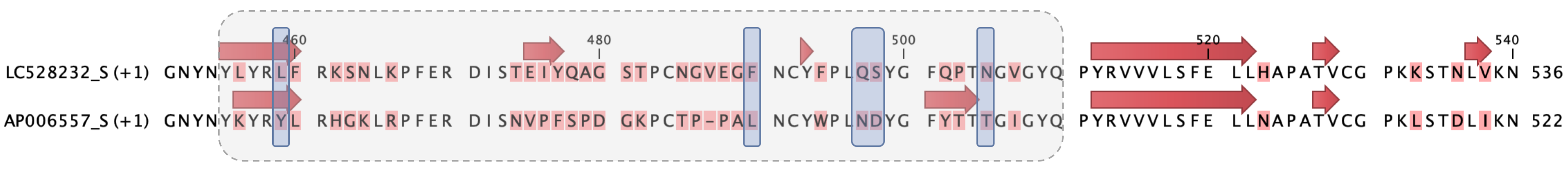}
    \caption{Five key mutations are shown using the blue boxes. These mutations are reported to change the binding affinity of the surface glycoprotein to the human ACE2 receptors. Comparisons are displayed for SARS-CoV-2 (LC528232) and SARS-CoV (AP006557).}
\label{fig:mutations}
\end{figure}

Wan and co-workers reported that mutations in the surface glycoprotein's five amino acids can play a critical role in binding to the human ACE2 receptors \cite{Wane0012720}. Anderson and co-workers provide their own insights, including another 6th critical amino acid. Here, we integrated secondary structure predictions and realigned the sequences in order to show how these five mutations may affect the binding of the SARS-CoV-2 protein (Figure \ref{fig:mutations}). The alignment of secondary structure predictions show the additions of three beta strands and the loss of one beta strand. These changes may affect the folding of the protein and hence its binding.

\begin{figure}[!thb]
\centering
  \includegraphics[width=1\textwidth]{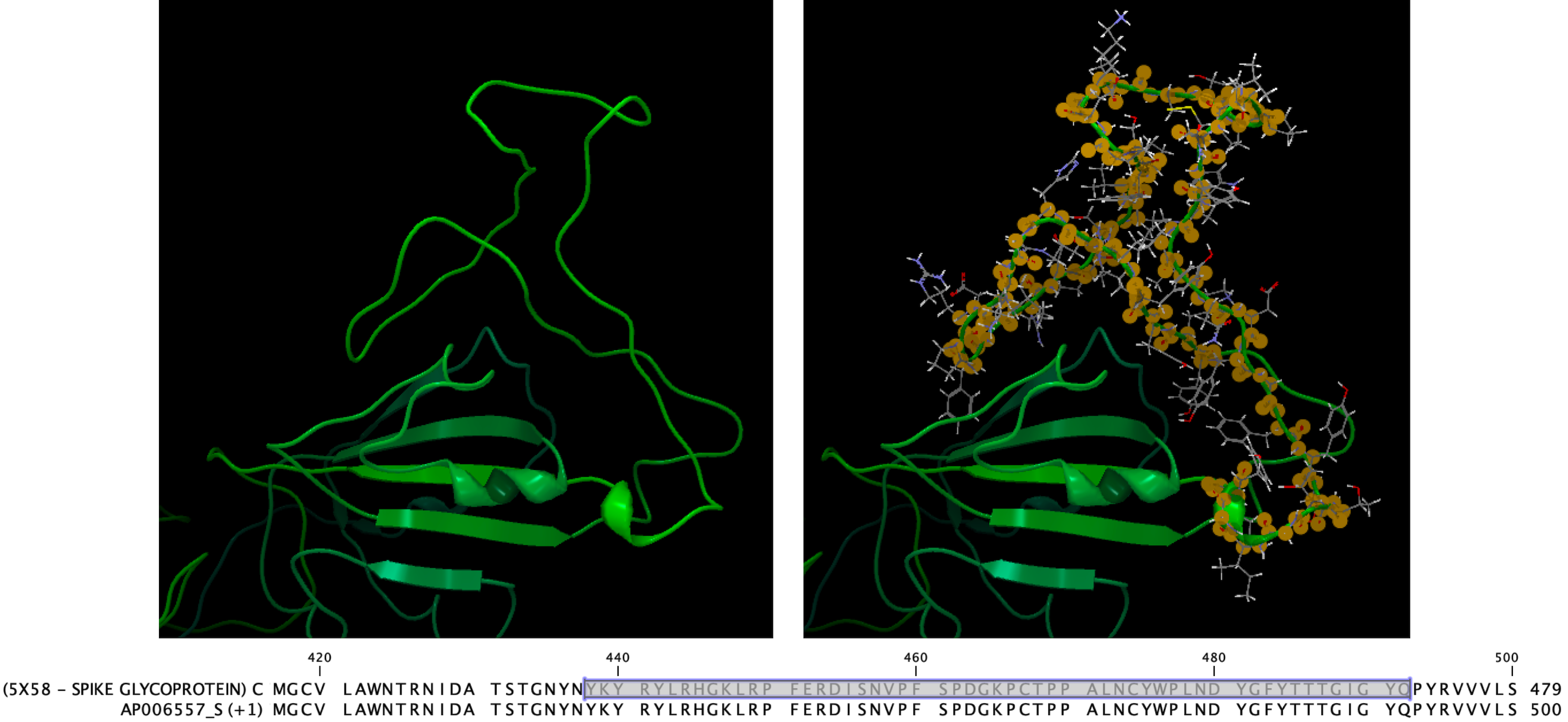}
    \caption{The binding region in SARS-CoV. The left picture shows the SARS-CoV Chain C, which binds to the human ACE2 receptors.  The right picture shows the area corresponding to the highlighted sequence below, which represent the mutation area shown using the dashed box in Figure \ref{fig:mutations}.}
\label{fig:sars-cov-mutation-region}
\end{figure}

Predictions for secondary structures may reveal some information about the binding of the surface glycoprotein. However, 3D models may help understanding the surfaces that are exposed and are likely to bind to other molecules. In order to understand the effects of the mutated region in Figure \ref{fig:mutations} (shown using the dashed box), an existing 3D model of the SARS-CoV protein was searched for. CLC Genomics Workbench's the `\texttt{Toolbox - Classical Sequence Analysis - Protein Analysis - Find and Model Structure}' option was used to search for existing 3D models. The first ranked entry with the most `Match identity' and `Coverage' was selected. The Protein Data Bank (PDB) \cite{Berman2003} identifier of this entry is `5X58 Prefusion Structure of SARS-CoV Spike Glycoprotein , Conformation 1' \cite{Yuan2017}. Using the CLC Genomics Workbench's `\texttt{Project Settings - Sequence tools - Show Sequence}' option, the 5X58's `Chain C' sequence was first displayed and it was then aligned to the SARS-CoV protein sequence (AP006557) using the `\texttt{Align to Existing Sequence}' option. The first picture on the left in Figure \ref{fig:sars-cov-mutation-region} shows the structure of the chain. The SARS-CoV amino acid sequences from the area with the key mutations (shown using the dashed box in Figure \ref{fig:mutations}) were highlighted using the sequence editor. The second picture on the right in Figure \ref{fig:sars-cov-mutation-region} shows the corresponding amino acid areas highlighted in the 3D structural view.

\begin{figure}[!thb]
\centering
  \includegraphics[width=1\textwidth]{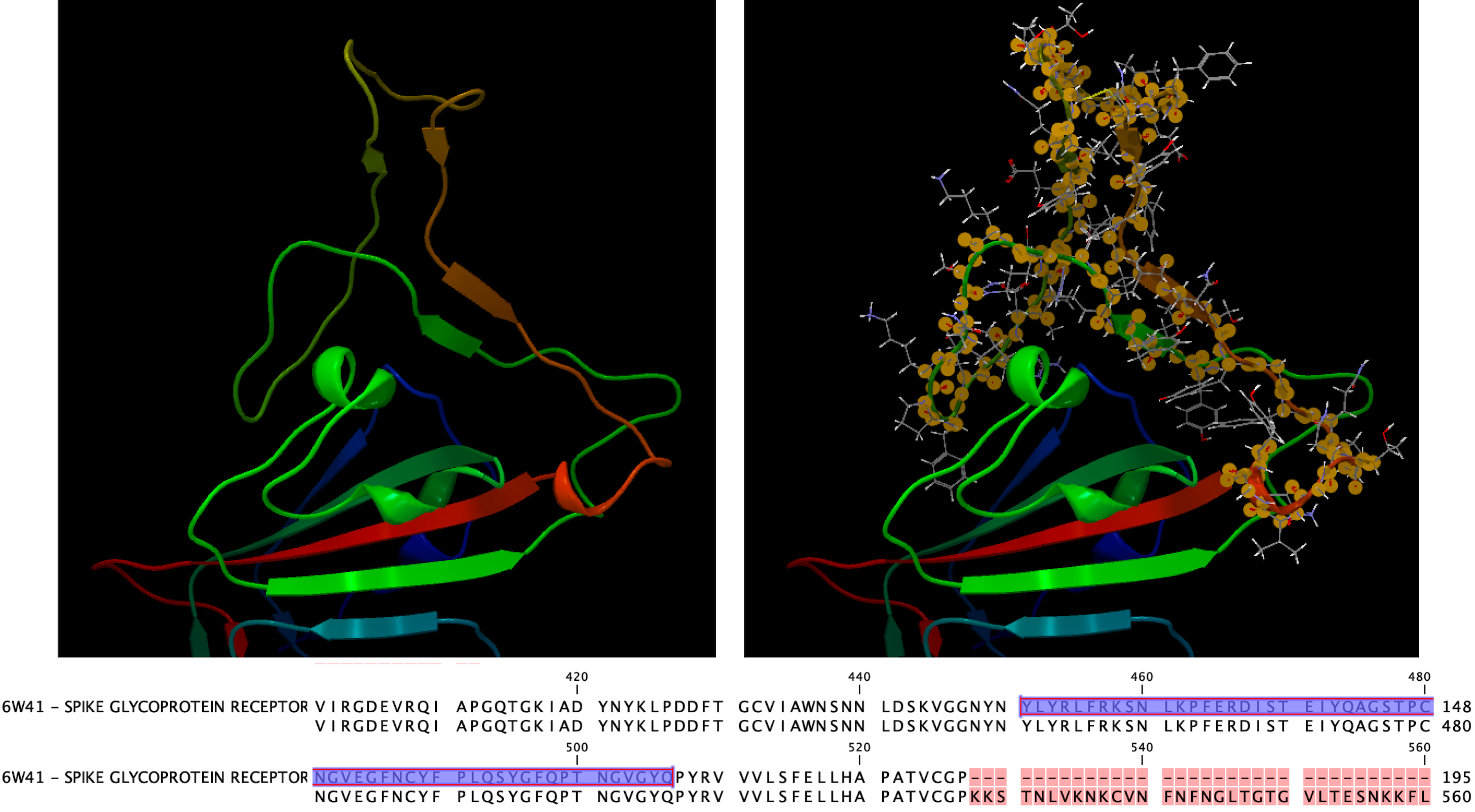}
    \caption{The binding region in SARS-CoV-2. The sequences from 6W41 and LC528232 were aligned. The left picture shows the SARS-CoV-2 Chain C, which binds to the human ACE2 receptors.  The right picture shows the area corresponding to the highlighted sequence below, which represents the mutation area shown using the dashed box in Figure \ref{fig:mutations}.}
\label{fig:sars-cov-2-mutation-region}
\end{figure}

A similar analysis was also carried out for the SARS-CoV-2 Chain C. The `6W41' entry \cite{6VW1}, including the details about the crystal structure of the SARS-CoV-2 receptor binding domain, was downloaded from the Protein Data Bank. A collection of related-entries can be found at the European Bioinformatics Institute's COVID-19 page \cite{EBI-COVID19}. The sequence from the `6W41' model was then aligned to the SARS-CoV-2 sequence (LC528232). Compared to the SARS-CoV secondary structures, both the CLC Genomics Workbench predictions and the SARS-CoV-2 `6W41' model reveal additional beta strands in the mutated region (shown using the dashed box in Figure \ref{fig:mutations}) of the surface glycoprotein. The 3D view of the binding region is shown in Figure \ref{fig:sars-cov-2-mutation-region}. The second picture on the right in Figure \ref{fig:sars-cov-2-mutation-region} shows the corresponding amino acid areas highlighted in the 3D structural view.

\begin{figure}[!thb]
\centering
  \includegraphics[width=1\textwidth]{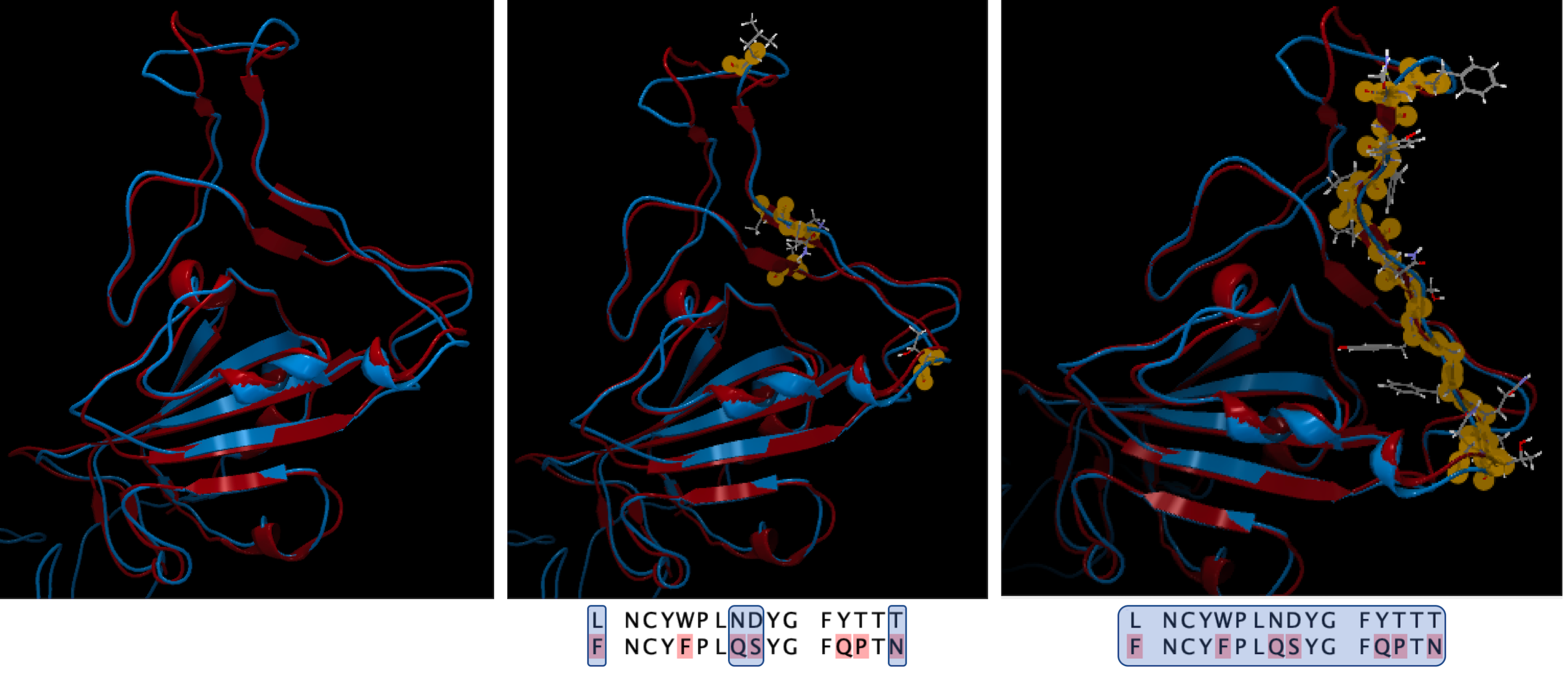}
    \caption{SARS-CoV-2 (in red colour) and SARS-CoV (in blue colour) 3D models of the binding regions to ACE2 sites are aligned. The middle picture shows the locations of the four key mutations in SARS-CoV-2. The third picture on the right shows the ACE2 binding area. SARS-CoV-2 sequences are shown at the bottom and the highlighted sequences are displayed using blue boxes.}
\label{fig:structure_alignment}
\end{figure}

CLC Genomic Workbench was then used to align the two SARS-CoV-2 and SARS-CoV 3D structures using the `\texttt{Project Settings - Structure tools - Align Protein Structure}' option.
Figure \ref{fig:structure_alignment} shows the alignment using red and blue colours for SARS-CoV-2 and SARS-CoV respectively. The third picture on the right shows the region that aligns with the ACE2 receptors during binding \cite{Shang2020}. The middle picture shows the four key amino acid sequences that mutated in SARS-CoV-2. These key mutations affect the binding affinity \cite{Shang2020}.

\section{Discussion}
In this report, insights from the sequence analysis of SARS-CoV-2 were presented. Genome-wide comparisons between SARS-CoV-2 and SARS-CoV further shed lights into understanding the potential effects of key mutations in the surface glycoprotein amino acid sequences. Our analysis is inline with the current reports \cite{Wu2020, Shang2020}. Although, some of the mutations may play an important role in binding to the human ACE2 receptors, there are several mutations that may have strengthened the binding affinity of the surface glycoprotein. Both the predictions and the 3D models reveal additional beta strands and hydrogen bonds. These additional mutations may cause changes in structural conformations and cause a higher binding affinity.

This report has been prepared in a tutorial style using CLC Genomics Workbench. The approach can be adopted by biologists and others to have initial understanding of SARS-CoV-2 mutations, and their relationships to SARS-CoV and other closely related viruses. Initial findings can then be explored further by using other specialised tools and approaches.

Computational methods are especially promising. Machine learning, artificial intelligence, data integration and mining, visualisation, computational and mathematical modelling for key biochemical interactions and disease control mechanisms can usefully be applied to provide solutions in a cost- and time- effective manner. We hope that this report is useful to those who wish to understand essential information about SARS-CoV-2 for subsequent analyses.

\section{Acknowledgements}
We thank QIAGEN for providing two months extended license beyond the two-week trial licence in order to utilise the CLC Genomics Workbench tool. The extended license was advertised by QIAGEN and was subsequently requested by the author.




\bibliography{references}

\providecommand{\latin}[1]{#1}
\makeatletter
\providecommand{\doi}
  {\begingroup\let\do\@makeother\dospecials
  \catcode`\{=1 \catcode`\}=2 \doi@aux}
\providecommand{\doi@aux}[1]{\endgroup\texttt{#1}}
\makeatother
\providecommand*\mcitethebibliography{\thebibliography}
\csname @ifundefined\endcsname{endmcitethebibliography}
  {\let\endmcitethebibliography\endthebibliography}{}
\begin{mcitethebibliography}{12}
\providecommand*\natexlab[1]{#1}
\providecommand*\mciteSetBstSublistMode[1]{}
\providecommand*\mciteSetBstMaxWidthForm[2]{}
\providecommand*\mciteBstWouldAddEndPuncttrue
  {\def\EndOfBibitem{\unskip.}}
\providecommand*\mciteBstWouldAddEndPunctfalse
  {\let\EndOfBibitem\relax}
\providecommand*\mciteSetBstMidEndSepPunct[3]{}
\providecommand*\mciteSetBstSublistLabelBeginEnd[3]{}
\providecommand*\EndOfBibitem{}
\mciteSetBstSublistMode{f}
\mciteSetBstMaxWidthForm{subitem}{(\alph{mcitesubitemcount})}
\mciteSetBstSublistLabelBeginEnd
  {\mcitemaxwidthsubitemform\space}
  {\relax}
  {\relax}

\bibitem[Zhou \latin{et~al.}(2020)Zhou, Yang, Wang, Hu, Zhang, Zhang, Si, Zhu,
  Li, Huang, Chen, Chen, Luo, Guo, Jiang, Liu, Chen, Shen, Wang, Zheng, Zhao,
  Chen, Deng, Liu, Yan, Zhan, Wang, Xiao, and Shi]{Zhou2020}
Zhou,~P.; Yang,~X.-L.; Wang,~X.-G.; Hu,~B.; Zhang,~L.; Zhang,~W.; Si,~H.-R.;
  Zhu,~Y.; Li,~B.; Huang,~C.-L.; Chen,~H.-D.; Chen,~J.; Luo,~Y.; Guo,~H.;
  Jiang,~R.-D.; Liu,~M.-Q.; Chen,~Y.; Shen,~X.-R.; Wang,~X.; Zheng,~X.-S.;
  Zhao,~K.; Chen,~Q.-J.; Deng,~F.; Liu,~L.-L.; Yan,~B.; Zhan,~F.-X.;
  Wang,~Y.-Y.; Xiao,~G.-F.; Shi,~Z.-L. {A pneumonia outbreak associated with a
  new coronavirus of probable bat origin}. \emph{Nature} \textbf{2020},
  \emph{579}, 270--273\relax
\mciteBstWouldAddEndPuncttrue
\mciteSetBstMidEndSepPunct{\mcitedefaultmidpunct}
{\mcitedefaultendpunct}{\mcitedefaultseppunct}\relax
\EndOfBibitem
\bibitem[NCB()]{NCBI-Taxonomy}
The National Center for Biotechnology, Taxonomy Browser.
  \url{https://www.ncbi.nlm.nih.gov/Taxonomy/Browser/wwwtax.cgi}\relax
\mciteBstWouldAddEndPuncttrue
\mciteSetBstMidEndSepPunct{\mcitedefaultmidpunct}
{\mcitedefaultendpunct}{\mcitedefaultseppunct}\relax
\EndOfBibitem
\bibitem[NCB()]{NCBI-Virus}
The National Center for Biotechnology, Virus Database.
  \url{https://www.ncbi.nlm.nih.gov/labs/virus/vssi}\relax
\mciteBstWouldAddEndPuncttrue
\mciteSetBstMidEndSepPunct{\mcitedefaultmidpunct}
{\mcitedefaultendpunct}{\mcitedefaultseppunct}\relax
\EndOfBibitem
\bibitem[Andersen \latin{et~al.}(2020)Andersen, Rambaut, Lipkin, Holmes, and
  Garry]{Andersen2020}
Andersen,~K.~G.; Rambaut,~A.; Lipkin,~W.~I.; Holmes,~E.~C.; Garry,~R.~F. {The
  proximal origin of SARS-CoV-2}. \emph{Nature Medicine} \textbf{2020}, \relax
\mciteBstWouldAddEndPunctfalse
\mciteSetBstMidEndSepPunct{\mcitedefaultmidpunct}
{}{\mcitedefaultseppunct}\relax
\EndOfBibitem
\bibitem[Wan \latin{et~al.}(2020)Wan, Shang, Graham, Baric, and
  Li]{Wane0012720}
Wan,~Y.; Shang,~J.; Graham,~R.; Baric,~R.~S.; Li,~F. Receptor Recognition by
  the Novel Coronavirus from Wuhan: an Analysis Based on Decade-Long Structural
  Studies of SARS Coronavirus. \emph{Journal of Virology} \textbf{2020},
  \emph{94}\relax
\mciteBstWouldAddEndPuncttrue
\mciteSetBstMidEndSepPunct{\mcitedefaultmidpunct}
{\mcitedefaultendpunct}{\mcitedefaultseppunct}\relax
\EndOfBibitem
\bibitem[Berman \latin{et~al.}(2003)Berman, Henrick, and Nakamura]{Berman2003}
Berman,~H.; Henrick,~K.; Nakamura,~H. {Announcing the worldwide Protein Data
  Bank}. \emph{Nature Structural {\&} Molecular Biology} \textbf{2003},
  \emph{10}, 980\relax
\mciteBstWouldAddEndPuncttrue
\mciteSetBstMidEndSepPunct{\mcitedefaultmidpunct}
{\mcitedefaultendpunct}{\mcitedefaultseppunct}\relax
\EndOfBibitem
\bibitem[Yuan \latin{et~al.}(2017)Yuan, Cao, Zhang, Ma, Qi, Wang, Lu, Wu, Yan,
  Shi, Zhang, and Gao]{Yuan2017}
Yuan,~Y.; Cao,~D.; Zhang,~Y.; Ma,~J.; Qi,~J.; Wang,~Q.; Lu,~G.; Wu,~Y.;
  Yan,~J.; Shi,~Y.; Zhang,~X.; Gao,~G.~F. {Cryo-EM structures of MERS-CoV and
  SARS-CoV spike glycoproteins reveal the dynamic receptor binding domains}.
  \emph{Nature Communications} \textbf{2017}, \emph{8}, 15092\relax
\mciteBstWouldAddEndPuncttrue
\mciteSetBstMidEndSepPunct{\mcitedefaultmidpunct}
{\mcitedefaultendpunct}{\mcitedefaultseppunct}\relax
\EndOfBibitem
\bibitem[6VW()]{6VW1}
Protein Data Bank, Structure of 2019-nCoV chimeric receptor-binding domain
  complexed with its receptor human ACE2.
  \url{http://www.rcsb.org/structure/6VW1}\relax
\mciteBstWouldAddEndPuncttrue
\mciteSetBstMidEndSepPunct{\mcitedefaultmidpunct}
{\mcitedefaultendpunct}{\mcitedefaultseppunct}\relax
\EndOfBibitem
\bibitem[EBI()]{EBI-COVID19}
The European Bioinformatics Institute, COVID-19 Outbreak.
  \url{https://www.ebi.ac.uk/ena/pathogens/covid-19}\relax
\mciteBstWouldAddEndPuncttrue
\mciteSetBstMidEndSepPunct{\mcitedefaultmidpunct}
{\mcitedefaultendpunct}{\mcitedefaultseppunct}\relax
\EndOfBibitem
\bibitem[Shang \latin{et~al.}(2020)Shang, Ye, Shi, Wan, Luo, Aihara, Geng,
  Auerbach, and Li]{Shang2020}
Shang,~J.; Ye,~G.; Shi,~K.; Wan,~Y.; Luo,~C.; Aihara,~H.; Geng,~Q.;
  Auerbach,~A.; Li,~F. {Structural basis of receptor recognition by
  SARS-CoV-2}. \emph{Nature} \textbf{2020}, \relax
\mciteBstWouldAddEndPunctfalse
\mciteSetBstMidEndSepPunct{\mcitedefaultmidpunct}
{}{\mcitedefaultseppunct}\relax
\EndOfBibitem
\bibitem[Wu \latin{et~al.}(2020)Wu, Zhao, Yu, Chen, Wang, Song, Hu, Tao, Tian,
  Pei, Yuan, Zhang, Dai, Liu, Wang, Zheng, Xu, Holmes, and Zhang]{Wu2020}
Wu,~F.; Zhao,~S.; Yu,~B.; Chen,~Y.-M.; Wang,~W.; Song,~Z.-G.; Hu,~Y.;
  Tao,~Z.-W.; Tian,~J.-H.; Pei,~Y.-Y.; Yuan,~M.-L.; Zhang,~Y.-L.; Dai,~F.-H.;
  Liu,~Y.; Wang,~Q.-M.; Zheng,~J.-J.; Xu,~L.; Holmes,~E.~C.; Zhang,~Y.-Z. {A
  new coronavirus associated with human respiratory disease in China}.
  \emph{Nature} \textbf{2020}, \emph{579}, 265--269\relax
\mciteBstWouldAddEndPuncttrue
\mciteSetBstMidEndSepPunct{\mcitedefaultmidpunct}
{\mcitedefaultendpunct}{\mcitedefaultseppunct}\relax
\EndOfBibitem
\end{mcitethebibliography}

\end{document}